\documentclass[conference]{IEEEtran}
\IEEEoverridecommandlockouts

\usepackage{cite}
\usepackage{amsmath,amssymb,amsfonts}
\usepackage{graphicx}
\usepackage{textcomp}
\usepackage{xcolor}
\usepackage{color}
\usepackage{booktabs}
\usepackage{multirow}
\usepackage{siunitx}
\usepackage{pgfplots}
\usepackage{pgfplotstable}
\usepackage{cite}
\usepackage{algorithm}
\usepackage{algorithmic}

\usepackage{colortbl}

\usetikzlibrary{fit}
\usepgfplotslibrary{groupplots}

\usepackage{array}
\usepackage{url}
\pgfplotsset{compat=1.18}
\usepackage{color,soul}
\usepackage{lipsum}  
\usepackage{multicol}

\usepackage{amsfonts}

\usepackage{tikzpeople}
\usepackage{epstopdf}

\usepackage{float}
\usepackage{tikz}
\usetikzlibrary{calc,positioning,shapes.geometric,shapes.symbols,shapes.misc}
\tikzstyle{decision} = [diamond, draw, text width=4.5em, text badly centered, node distance=3cm, inner sep=0pt]

\usepackage{eurosym}
\usetikzlibrary{shapes.geometric, arrows.meta, positioning}

\usepackage{amssymb}

\usepackage{tabularx,booktabs}

\begin{document}

\title{Droop-Aware Foundation Model Power Flow
\thanks{This work relates to the Department of Navy award N000142512374 issued by the Office of Naval Research. \textit{Corresponding Author: Peng Zhang.}}
}

\author{\IEEEauthorblockN{Khezr Sanjani and Peng Zhang}
\thanks{
         K. Sanjani and P. Zhang are with the Department of Electrical and Computer Engineering, Stony Brook University, Stony Brook, NY, 11794-2350, USA.  (e-mail: khezr.sanjani, P.Zhang@stonybrook.edu). 
         } 

}

\maketitle


\begin{abstract}
This paper develops a droop-aware extension of the GridFM power systems foundation model, embedding droop gains and frequency/voltage deadband parameters as per-bus node features to enable control-aware AC power-flow analysis. Existing power-flow datasets encode only static electrical features, conflating operating points from qualitatively different control regimes; this work resolves that gap by exposing droop and deadband parameters as structured node features, with deadband discontinuities handled through a smooth $\tanh$ approximation that preserves solver differentiability. A transformer-based graph neural network is pre-trained on masked reconstruction and fine-tuned on the resulting control-aware datasets. The framework is validated against PSCAD electromagnetic-transient simulations on a two-bus system ($0.11\%$ maximum steady-state error) and cross-validated against an independent PyPower droop solver on the IEEE 24-bus RTS. 
On the 24-bus system the surrogate attains $R^2 = 0.9996$ for active generation and $0.0015$~p.u.\ voltage-magnitude RMSE; scalability is confirmed on the IEEE 300-bus system ($0.0036$~p.u.\ RMSE, $R^2 = 0.9841$ for voltage magnitude across $299{,}700$ predictions). A three-mode control study further shows that the deadband widens the control-error distribution while leaving total droop compensation unchanged, establishing deadband width as an actionable node-level design feature.
\end{abstract}


\section{Introduction}
The transition toward inverter-dominated power systems has reshaped the steady-state behavior of transmission and distribution networks. Grid-forming (GFM) inverters, which impose local voltage and frequency references, have become a key enabler of high-penetration renewable integration~\cite{feng2022distributed}, and among GFM strategies, droop control remains the most widely deployed owing to its decentralized structure and its emulation of the load-sharing behavior of synchronous machines~\cite{9099877}. Accurate representation of droop gains and deadband thresholds is therefore essential for realistic steady-state analysis of inverter-rich grids~\cite{vorobev2019deadbands}.

In parallel, machine-learning surrogates have gained traction as computationally efficient alternatives to repeated numerical solves. Graph neural networks (GNNs) are particularly well suited to power systems because they respect the graph-structured topology of transmission networks and propagate information through message passing along physical edges~\cite{Bottcher2022}. Prior work has demonstrated GNN surrogates for AC power-flow prediction~\cite{Bottcher2022}, optimal power-flow approximation~\cite{owerko2020opf}, and operational risk quantification~\cite{zhang2024risk} across systems of varying scale. Foundation models (FMs) extend this paradigm: trained on large-scale, diverse datasets, they learn the underlying physics, topology, and operating behavior of power networks across a wide range of configurations. GridFM pioneers and exemplifies this direction, encoding the transmission network as a graph and training a \textit{GraphTransformerConv} through masked node reconstruction with a physics-informed loss that enforces nodal power balance~\cite{puech2025gridfm}. Unlike task-specific surrogates that require retraining for every new operating condition, a pre-trained FM generalizes across configurations and can be fine-tuned efficiently for downstream tasks~\cite{hamann2024gridfm}.

However, existing FM-based and GNN-based frameworks encode only static electrical features---load demand, generation dispatch, and network impedances---leaving droop control parameters unrepresented. To the best of the authors' knowledge, no prior surrogate uses droop gains and deadband thresholds as per-bus node features for steady-state power-flow prediction: existing GNN surrogates parameterize operating points by load and topology alone~\cite{Bottcher2022, owerko2020opf}, while droop-aware power-flow formulations remain confined to numerical solvers~\cite{feng2022distributed, mumtaz2016droop}. Under droop control, the steady-state operating point is determined not by load and topology alone but also by the droop gain and deadband of each generator bus~\cite{feng2022distributed}: two systems with identical loads but different droop settings converge to different voltage and power dispatches. A dataset that omits these parameters therefore conflates operating points from fundamentally different control configurations~\cite{9099877}, degrading model accuracy and preventing meaningful scenario analysis across control settings~\cite{hamann2024gridfm}. This limitation is compounded by the coupling introduced by droop control, where the shared frequency deviation creates long-range dependencies between buses that motivate an attention-based graph architecture~\cite{mumtaz2016droop}.

To address this gap, this paper extends the GridFM framework with practical droop control and associated deadband modeling as per-bus node features. Because the resulting model is pre-trained across diverse control configurations and fine-tuned as a control-aware surrogate---rather than retrained per operating condition---it retains the transferable, reusable character that motivates the foundation-model paradigm~\cite{hamann2024gridfm}. The contributions are as follows: 

\begin{enumerate}
    \item Active/reactive power droop gains and frequency/voltage deadband thresholds are implemented within the GridFM framework as structured bus-level node features, with deadband discontinuities resolved via a smooth $\tanh$ approximation that preserves solver differentiability, enabling control-aware dataset generation.
    \item The modified framework is validated against electromagnetic-transient (EMT) simulations on a two-bus PSCAD reference model and cross-validated against an independent PyPower droop solver on the IEEE 24-bus RTS, confirming the correctness of the droop/deadband implementation. 
    \item A transformer-based GNN trained on the droop-augmented datasets attains high steady-state accuracy on the IEEE 24-bus RTS and scales to the IEEE 300-bus system, verifying accuracy and feasibility at scale.
    \item A three-mode control study (slack-bus, droop-only, and droop with deadband) quantifies the behavioral contribution of the deadband, establishing deadband width as an actionable node-level design feature.
\end{enumerate}

\section{Methodology}
\label{sec:method}
This section presents the control-aware framework in five parts. Section~\ref{sec:data} describes the data-generation pipeline that extends \textit{gridfm-datakit} with per-bus droop constraints. Section~\ref{sec:droop} formulates the $P$--$f$ and $Q$--$V$ droop relations and their coupling to the AC power-flow problem, and Section~\ref{sec:deadband} introduces the smooth $\tanh$ deadband approximation that preserves solver differentiability. Section~\ref{sec:fm} details the two-stage pre-training and fine-tuning workflow together with the masking and inference protocol, and Section~\ref{sec:gnn} defines the augmented node feature vector that embeds droop gains and deadband thresholds alongside standard electrical quantities. Together, these components produce a surrogate whose inputs span both the electrical state and the control configuration of the network, enabling scenario analysis across droop settings without retraining.

\subsection{Dataset Generation}
\label{sec:data}
The control-aware data-generation pipeline extends the \textit{gridfm-datakit} perturbation workflow~\cite{puech2025gridfm} by enforcing per-bus droop constraints, producing operating points across diverse control configurations. Load scenarios are generated using a hybrid perturbation strategy that combines a global temporal scaling factor derived from aggregated real-world load profiles with independent per-bus noise. At each time step $t$, the perturbed active and reactive demands are 

\begin{equation}
\tilde{p}_{i,t} = p_i \cdot \mathrm{ref}_t \cdot \epsilon^{p}_{i,t},
\qquad
\tilde{q}_{i,t} = q_i \cdot \mathrm{ref}_t \cdot \epsilon^{q}_{i,t},
\label{eq:loadpert}
\end{equation}

where $\epsilon^{p}_{i,t}, \epsilon^{q}_{i,t} \sim \mathcal{U}(1-\sigma,\,1+\sigma)$ introduce per-bus variability around the global factor $\mathrm{ref}_t$. This preserves the spatial correlation and temporal realism of the aggregate profile while retaining sufficient scenario diversity. Network topology and branch admittances are held fixed across scenarios, so that variation arises from load and control configuration alone.

The global scaling factor is selected by a feasibility search: the base AC optimal power flow (ACOPF) is solved at increasing load levels until the largest factor admitting a converged solution is identified, and scenarios are then drawn from a band below this ceiling. Generator setpoints $\{P^{\mathrm{set}}_{g,i}, Q^{\mathrm{set}}_{g,i}, V_{0,i}\}$ are initialized from this base ACOPF solution, after which the droop relations of Section~\ref{sec:droop} are imposed as additional equality constraints and the coupled system is solved to yield the final control-aware operating point. Because droop power flow holds generator setpoints fixed rather than re-dispatching, the admissible load band is narrower than for ACOPF: excessive deviation drives generators into their power limits and renders the problem infeasible.


\subsection{Droop Control Formulation}
\label{sec:droop}
Without loss of generality, this subsection formulates the $P-f$ and $Q-V$ droop relations that govern generator bus injections, replacing the fixed-injection assumption of standard AC power flow with a control-parameterized model that directly couples network state to droop gain. Let $\mathcal{D} \subseteq \{1,\ldots,N\}$ denote the set of droop-participating generator buses. Their active/reactive injections are governed by 
\begin{equation} \vspace{-9pt}
  P_{g,i} = P_{g,i}^{\mathrm{set}} - \frac{1}{m_{p,i}}\,\Delta f_{\mathrm{eff}},
  \label{eq:pf_droop} 
\end{equation}\vspace{4pt}
\begin{equation} 
  Q_{g,i} = Q_{g,i}^{\mathrm{set}} - \frac{1}{m_{q,i}}\!\left(V_i - V_{0,i}\right),
  \label{eq:qv_droop}  
\end{equation}
where $m_{p,i}$/$m_{q,i}$ are the per-unit droop gains, $V_{0,i}$ is the voltage reference, and $\Delta f_{\mathrm{eff}}$ is the deadband frequency deviation defined in Section \ref{sec:deadband}. Droop/deadband parameters are randomized to produce a control-parameterized dataset.

\subsection{Deadband Implementation}
\label{sec:deadband}
This subsection introduces a smooth tanh approximation of the hard frequency deadband~\cite{vorobev2019deadbands}, resolving the derivative discontinuity that impedes convergence of the interior-point NLP solver and enabling deadband width to be treated as a differentiable, trainable node feature.
A hard frequency deadband, inactive for $|\Delta f| \leq db$ and active otherwise, introduces a discontinuity in the constraint Jacobian, causing solver stagnation near $\Delta f = 0$. To preserve differentiability, the hard threshold is replaced by a smooth blending function:

\begin{equation} \vspace{-10pt}
  \Delta f_{\mathrm{eff}}
  = \Delta f \Bigl[\epsilon + (1-\epsilon)\,a(\Delta f)\Bigr],
  \label{eq:deadband}\vspace{8pt}
\end{equation}
where $\epsilon = 0.05$ retains a $5\%$ residual droop response in the deadband to prevent stagnation, and the activation function is:
\begin{equation}\vspace{-4pt}
  a(\Delta f)
  = \frac{1}{2}
  \left[1 + \tanh\!\left(\sigma\,\frac{|\Delta f|-db}{db+\delta}\right)\right],
  \label{eq:activation} 
\end{equation}
For instance, 
deadband half-width $db = 0.006\,\text{p.u.}$, steepness $\sigma = 10$, and regularization $\delta = 10^{-6}\,\text{p.u.}$ As $\epsilon \to 0$ with large $\sigma$, \eqref{eq:deadband} recovers the ideal piecewise deadband characteristic.

Although $a(\Delta f)$ contains $|\Delta f|$ and is therefore not
differentiable at $\Delta f = 0$, the effective deviation
$\Delta f_{\mathrm{eff}}$ is. Differentiating~\eqref{eq:deadband},

\begin{equation}
\frac{\partial \Delta f_{\mathrm{eff}}}{\partial \Delta f}
= \bigl[\epsilon + (1-\epsilon)\,a(\Delta f)\bigr]
+ \Delta f\,(1-\epsilon)\,a'(\Delta f),
\end{equation}

the discontinuous term $a'(\Delta f)$ is multiplied by $\Delta f$,
which vanishes precisely at $\Delta f = 0$; hence
$\partial \Delta f_{\mathrm{eff}}/\partial \Delta f$ remains
continuous there. The residual non-smoothness is confined to the
second derivative. The smooth deadband therefore removes the
first-order derivative discontinuity responsible for solver
stagnation near $\Delta f = 0$.

\subsection{Foundation Model Training}
\label{sec:fm}
Here, we describe the training workflow, in which the GridFM backbone is trained on masked node reconstruction with a physics-informed loss over the droop-augmented dataset. In each epoch, $50\%$ of the maskable node features are randomly masked and the model minimizes a joint loss:
\begin{equation}\vspace{-5pt}
    \mathcal{L} = \mathcal{L}_{\mathrm{rec}} + \lambda \mathcal{L}_{\mathrm{phys}},
    \label{eq:loss}
\end{equation}
where $\mathcal{L}_{\mathrm{rec}}$ is the masked node reconstruction loss and $\mathcal{L}_{\mathrm{phys}}$ penalizes violations of nodal power balance, embedding network physics directly into the learned representations \cite{puech2025gridfm}. Optimization uses Adam at a learning rate of $10^{-4}$, with the physics term weighted more heavily than the reconstruction term. Because the backbone is trained across diverse control configurations rather than a single operating condition, it retains the transferability that motivates the foundation-model paradigm, and can be fine-tuned for unseen grid configurations with minimal labeled data \cite{hamann2024gridfm}.

At inference, the six nodal quantities
$\{P_d, Q_d, P_g, Q_g, V_m, V_a\}$ serve as both inputs and
reconstruction targets under a masked-prediction protocol: a subset
is masked and recovered from the remaining observed features. The
droop/deadband parameters $\{m_p, m_q, db^{f}, db^{v}\}$, the bus-type
indicators $\{t^{PQ}, t^{PV}, t^{REF}\}$, and all edge features are
always observed and are never masked. The held-out metrics reported
in Section~III are computed under the same $50\%$ random-masking
protocol used in training: for each held-out scenario the model
reconstructs the masked node features from the remaining observed
features together with the fixed control and topology features. The GNN architecture and training hyperparameters are summarized in
Table~\ref{tab:hyperparams}.

\begin{table}[h]
\centering
\caption{GNN Architecture and Training Hyperparameters}
\label{tab:hyperparams}
\begin{tabular}{ll}
\hline
Hyperparameter & Value \\
\hline
GNN layers $L$            & 4 \\
Hidden dimension $d$      & 128 \\
Attention heads           & 4 \\
Positional encoding dim.  & 20 \\
Node features $d_v$       & 13 \\
Edge features $d_e$       & 2 \\
Training epochs           & 100 \\
Batch size                & 128 \\
Optimiser                 & Adam ($\beta_1{=}0.9,\ \beta_2{=}0.999$) \\
Learning rate             & $1\times10^{-4}$ \\
LR scheduler              & ReduceLROnPlateau ($\gamma{=}0.7$, patience 5) \\
Masking ratio             & 50\% \\
Loss weights $[\mathcal{L}_{\mathrm{rec}}, \mathcal{L}_{\mathrm{phys}}]$ & $[0.1,\ 0.9]$ \\
Early stopping patience   & 15 \\
Random seed               & 42 \\
\hline
\end{tabular}
\end{table}

\subsection{GNN for Droop-Aware Power Flow Surrogate}
\label{sec:gnn}

This subsection defines the augmented node feature vector that embeds droop gains and deadband thresholds alongside standard electrical quantities, constituting the core architectural extension that makes the surrogate control-aware.
The transmission network is represented as an undirected graph 
$\mathcal{G} = (\mathcal{V}, \mathcal{E})$, where the node feature 
vector augments standard quantities with per-bus droop/deadband parameters:
\begin{multline}  
  \mathbf{x}_i = \bigl[
    P_{d,i},\; Q_{d,i},\; V_{m,i},\; \theta_i,\;
    P_{g,i},\; Q_{g,i},\; m_{p,i},\; m_{q,i},\\
    db^{f}_{i},\; db^{v}_{i},\;
    t_i^{\mathrm{PQ}},\; t_i^{\mathrm{PV}},\;
    t_i^{\mathrm{REF}}
  \bigr]^{\!\top} \in \mathbb{R}^{13},
  \label{eq:node_features}
\end{multline}
with droop entries set to zero for non-participating buses. Edge features are the real and imaginary parts of the branch admittance, $\mathbf{e}_{ij} = [G_{ij}, B_{ij}]^{\top} \in \mathbb{R}^{2}$.

The overall algorithm is summarized in Algorithm 1.

\begin{algorithm}[]
\caption{Droop-Aware Foundation Model Power-Flow}
\label{alg:full}
\begin{algorithmic}[1]
\STATE \textbf{Input:} Grid $\mathcal{G}=(\mathcal{V},\mathcal{E})$,\;
loads $\{p_i, q_i\}$,\; EIA profiles,\;
droop ranges,\; deadband $db$,\; noise $\sigma$,\; target size $S$

\vspace{3pt}
\STATE \textbf{Part I — Dataset Generation}
\FOR{each scenario $s = 1,\ldots,S$}
    \STATE Perturb loads;
           solve base ACOPF $\rightarrow$ setpoints
           $P^{\mathrm{set}}_{g,i}$, $Q^{\mathrm{set}}_{g,i}$, $V_{0,i}$;
           draw random droop gains $m_{p,i}$, $m_{q,i}$
    \STATE Enforce P-f and Q-V droop \eqref{eq:pf_droop}\eqref{eq:qv_droop},
           compute deadband-filtered $\Delta f_{\mathrm{eff}}$ \eqref{eq:deadband},
           solve coupled AC power-flow;
           \textbf{discard if infeasible}
    \STATE Store $\mathbf{x}_i \in \mathbb{R}^{13}$,\;
           $\mathbf{e}_{ij} \in \mathbb{R}^{5}$,\;
           targets $\mathbf{y}_i$
\ENDFOR

\vspace{3pt}
\STATE \textbf{Output:} $\mathcal{D}=\{(\mathbf{x}_i^{(s)},\mathbf{e}_{ij}^{(s)},\mathbf{y}_i^{(s)})\}_{s=1}^{S}$

\vspace{3pt}
\STATE \textbf{Part II — GNN Training and Inference}
\STATE Split $\mathcal{D}$: 90\% train / 10\% test
\STATE \textbf{Pre-train}: minimise
       $\mathcal{L} = \mathcal{L}_{\mathrm{rec}} + \lambda\,\mathcal{L}_{\mathrm{phys}}$
       with 30\% node masking \eqref{eq:loss}
\STATE \textbf{Fine-tune}: supervised MSE over
       $\{P_d, Q_d, P_g, Q_g, V_m, V_a\}$;\;
       Adam, lr\,$= 10^{-5}$;\;
       forward-pass held-out scenarios through $f_\theta$

\vspace{3pt}
\STATE \textbf{Output:} $\{\hat{P}_d, \hat{Q}_d, \hat{P}_g, \hat{Q}_g, \hat{V}_m, \hat{V}_a\}$;\;
evaluate $R^2$, RMSE, MAE, Bias
\end{algorithmic} 
\end{algorithm}

\vspace{5pt}
\section{Results and Analysis}
\label{sec:results}

The proposed FM extension model is evaluated on three systems including a two-bus EMT validation case, the IEEE 24-bus RTS, and the IEEE 300-bus system, implemented in Julia/PowerModels.jl.

\subsection{Validation Against EMT Results}
\label{sec:pscad}

\begin{figure}[h]
    \centering
    \begin{tikzpicture}
        \begin{axis}[
            width=\linewidth, height=5.5cm,
            xlabel={Time (s)},
            ylabel={Voltage (p.u.)},
            grid=major,
            grid style={line width=0.3pt, draw=gray!40},
            major grid style={line width=0.5pt, draw=gray!50},
            legend pos=south west,
            legend style={
                font=\small,
                draw=black!50,
                fill=white,
                fill opacity=0.9,
                text opacity=1,
                inner sep=3pt,
                row sep=1pt,
            },
            xmin=0, xmax=30,
            ymin=0.97, ymax=1.05,
            xtick={0,5,10,15,20,25,30},
            ytick={0.98,1.00,1.02,1.04},
            tick label style={font=\small},
            label style={font=\small},
            axis line style={line width=0.6pt},
            clip=false,
            /pgf/number format/1000 sep={},
        ]
            \addplot[
                color=blue!80!black,
                line width=1.0pt,
                mark=none,
                unbounded coords=jump,
                clip=true,
            ] table [
                x=time,
                y=PSCAD,
                col sep=comma,
                header=true,
            ] {PSCAD_GridFM_Validation.csv};
            \addlegendentry{$PSCAD$}

            \addplot[
                color=red!80!black,
                line width=1.4pt,
                dashed,
                mark=none,
                clip=true,
            ] table [
                x=time,
                y=GridFM,
                col sep=comma,
                header=true,
            ] {PSCAD_GridFM_Validation.csv};
            \addlegendentry{$GridFM$}

            \draw[densely dashed, gray!60, line width=0.6pt]
                (axis cs:10,0.97) -- (axis cs:10,1.052);
            \draw[densely dashed, gray!60, line width=0.6pt]
                (axis cs:20,0.97) -- (axis cs:20,1.052);


            \node[font=\tiny, text=blue!70!black, anchor=west]
                at (axis cs:3.0, 1.042) {$\mathrm{er}{=}0.01\%$};
            \draw[->, blue!70!black, line width=0.6pt]
                (axis cs:5.5, 1.041) -- (axis cs:5.5, 1.0285);

            \node[font=\tiny, text=blue!70!black, anchor=west]
                at (axis cs:12.0, 1.046) {$\mathrm{er}{=}0.10\%$};
            \draw[->, blue!70!black, line width=0.6pt]
                (axis cs:14.5, 1.045) -- (axis cs:14.5, 1.0351);

            \node[font=\tiny, text=blue!70!black, anchor=west]
                at (axis cs:24.0, 1.034) {$\mathrm{er}{=}0.11\%$};
            \draw[->, blue!70!black, line width=0.6pt]
                (axis cs:27.5, 1.033) -- (axis cs:27.5, 1.0164);

        \end{axis}
    \end{tikzpicture}\vspace{-12pt}
    \caption{Voltage at Bus~2 across three load scenarios.
        GridFM (dashed) represents the steady-state power flow
        prediction; PSCAD (solid) captures the full EMT response. Annotations show steady-state error
        per scenario.}
    \label{fig:2busvalidation}
\end{figure}
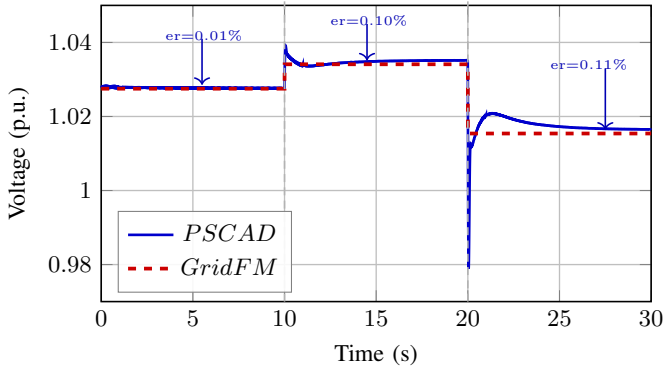

The droop/deadband implementation is validated against a PSCAD reference model on a two-bus system. As shown in Fig.~\ref{fig:2busvalidation}, three load scaling events are applied sequentially, with GridFM steady-state predictions tracking the PSCAD electromagnetic transient response at each operating point. The maximum bus-level discrepancy across all events is $0.11\%$, with the residual difference attributable to the switching-level detail in PSCAD versus the phasor-domain formulation in GridFM.

\begin{table}[h] \vspace{-10pt}
  \centering
  \caption{Frequency Cross-Validation: Julia/GridFM vs.\ PyPower-Droop} \vspace{-8pt}
  \label{tab:crossval}
  \renewcommand{\arraystretch}{1.15}
  \begin{tabular}{lccc}
    \toprule
    \textbf{Metric}              & \textbf{Julia/GridFM} & \textbf{PyPower} & \textbf{Diff.} \\
    \midrule
    $\Delta f$ (p.u.)            & $-$0.001673 & $-$0.001648 & 0.000025 \\
    System frequency (Hz)        & 59.8996     & 59.9011     & 0.0015   \\
    Relative error               & ---         & ---         & 1.51\%   \\
    \midrule
    Total generation (MW)        & 2904.47     & 2978.18     & ---      \\
    Total load (MW)              & 2867.86     & 2930.96     & ---      \\
    System losses (MW)           & 36.61       & 47.22       & $+$10.61 \\
    \bottomrule
  \end{tabular}
  \vspace{-12pt}
\end{table} 

\subsection{GNN Surrogate Accuracy: IEEE 24-Bus RTS}
\label{sec:24bus}

\begin{figure}[t] \vspace{-5pt}
\centering
\begin{tikzpicture}
\pgfplotsset{
    scatter style/.style={
        only marks,
        mark=*,
        mark size=0.6pt,
        color=blue!70!black,
        fill opacity=0.5,
    },
    diag style/.style={
        no marks,
        dashed,
        color=red!80!black,
        line width=0.9pt,
    },
}
\begin{groupplot}[
    group style={
        group size=3 by 2,
        horizontal sep=0.75cm,
        vertical sep=1.15cm,
    },
    width=3.6cm, height=3.6cm,
    tick label style={font=\tiny},
    label style={font=\tiny},
    title style={font=\footnotesize, at={(0.5,0.9)}, anchor=south},
    xlabel style={font=\tiny, yshift=2pt},
    ylabel style={font=\tiny},
    grid=major,
    grid style={line width=0.2pt, draw=gray!30},
    major grid style={line width=0.3pt, draw=gray!40},
    axis line style={line width=0.5pt},
    scaled ticks=false,
    max space between ticks=20,
    xlabel={True},
    ylabel={Pred},
]

\nextgroupplot[
    title={$P_d$ (MW)},
    xmin=-20, xmax=350, ymin=-20, ymax=350,
    xtick={0,150,300}, ytick={0,150,300},
]
\addplot[scatter style] table[x=Pd_t, y=Pd_p, col sep=comma]{scatter_data.csv};
\addplot[diag style] coordinates {(-20,-20)(350,350)};

\nextgroupplot[
    title={$Q_d$ (MVAr)},
    ylabel={},
    xmin=-15, xmax=80, ymin=-15, ymax=80,
    xtick={0,30,60}, ytick={0,30,60},
]
\addplot[scatter style] table[x=Qd_t, y=Qd_p, col sep=comma]{scatter_data.csv};
\addplot[diag style] coordinates {(-15,-15)(80,80)};

\nextgroupplot[
    title={$P_g$ (MW)},
    ylabel={},
    xmin=-20, xmax=700, ymin=-20, ymax=700,
    xtick={0,300,600}, ytick={0,300,600},
]
\addplot[scatter style] table[x=Pg_t, y=Pg_p, col sep=comma]{scatter_data.csv};
\addplot[diag style] coordinates {(-20,-20)(700,700)}; 

\nextgroupplot[
    title={$Q_g$ (MVAr)},
    xmin=-90, xmax=200, ymin=-90, ymax=200,
    xtick={-50,50,150}, ytick={-50,50,150},
]
\addplot[scatter style] table[x=Qg_t, y=Qg_p, col sep=comma]{scatter_data.csv};
\addplot[diag style] coordinates {(-90,-90)(200,200)};

\nextgroupplot[
    title={$V_m$ (p.u.)},
    ylabel={},
    xmin=1.05, xmax=1.16, ymin=1.05, ymax=1.16,
    xtick={1.07,1.11,1.15}, ytick={1.07,1.11,1.15},
    x tick label style={rotate=30, anchor=east, font=\tiny},
]
\addplot[scatter style] table[x=Vm_t, y=Vm_p, col sep=comma]{scatter_data.csv};
\addplot[diag style] coordinates {(1.05,1.05)(1.16,1.16)};

\nextgroupplot[
    title={$V_a$ (deg)},
    ylabel={},
    xmin=-6, xmax=25, ymin=-6, ymax=25,
    xtick={0,10,20}, ytick={0,10,20},
]
\addplot[scatter style] table[x=Va_t, y=Va_p, col sep=comma]{scatter_data.csv};
\addplot[diag style] coordinates {(-6,-6)(25,25)};

\end{groupplot}
\end{tikzpicture}\vspace{-10pt}
\caption{True vs.\ predicted ($P_d$, $Q_d$, $P_g$, $Q_g$, $V_m$, $V_a$) for the IEEE~24-bus. }
\label{fig:scatter} \vspace{-15pt}
\end{figure}
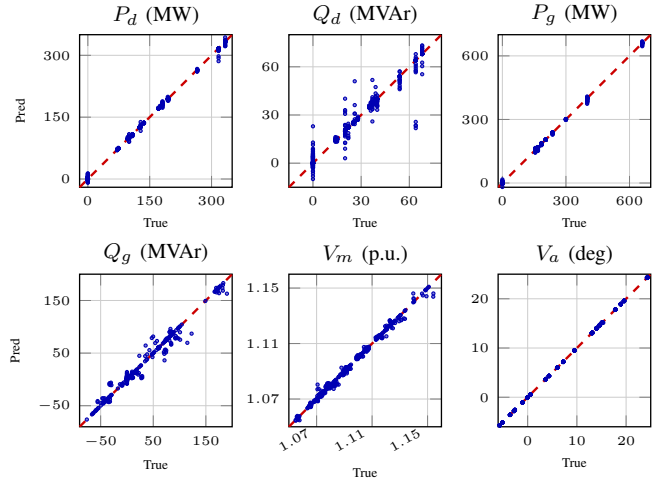

Notably, these results are achieved by a FM trained across diverse droop configurations rather than a task-specific surrogate trained on a fixed control setting. Fig.~\ref{fig:scatter} confirms tight clustering around the ideal diagonal for all six targets. Table~\ref{tab:gnn_24bus} reports performance across 1,000 held-out scenarios, with key observations:

\begin{itemize}
    \item $R^2 \geq 0.999$ is achieved for $P_g$ and $V_a$, with voltage magnitude recovered at RMSE~$= 0.0015\,$p.u. and angle error below $0.032^{\circ}$.
    \item Reactive quantities ($Q_d$, $Q_g$) show lower but still strong $R^2$ values of $0.940$ and $0.974$, consistent with the known sensitivity of reactive injections to the nonlinear $Q$--$V$ characteristic under varying droop gains.
    \item Despite the lower $R^2$ for reactive quantities, the absolute RMSE of $4.935$~MVAr for $Q_d$ and $7.743$~MVAr for $Q_g$ remain within operationally acceptable bounds for transmission-level dispatch.
\end{itemize}

\begin{table}[h] \vspace{-10pt}
  \centering
  \caption{GNN Performance on the IEEE 24-Bus RTS} \vspace{-8pt}
  \label{tab:gnn_24bus}
  \renewcommand{\arraystretch}{1.15}
  \begin{tabular}{lcccc}
    \toprule
    Target        & RMSE   & MAE    & Bias       & $R^{2}$  \\
    \midrule
    $P_d$ (MW)    & 3.567  & 1.448  & $-$0.262   & 0.9987   \\
    $Q_d$ (MVAr)  & 4.935  & 1.834  & $-$0.086   & 0.9403   \\
    $P_g$ (MW)    & 3.615  & 1.540  & $-$0.599   & 0.9996   \\
    $Q_g$ (MVAr)  & 7.743  & 3.122  & $-$0.286   & 0.9740   \\
    $V_m$ (p.u.)  & 0.0015 & 0.0008 & $-$0.00008 & 0.9951   \\
    $V_a$ (deg)   & 0.0311 & 0.0162 & $-$0.0010  & 0.99999  \\
    \bottomrule
  \end{tabular}
  \vspace{-5pt}

  {}
\end{table}

\begin{figure}[t]
  \centering
  \includegraphics[width=\columnwidth]{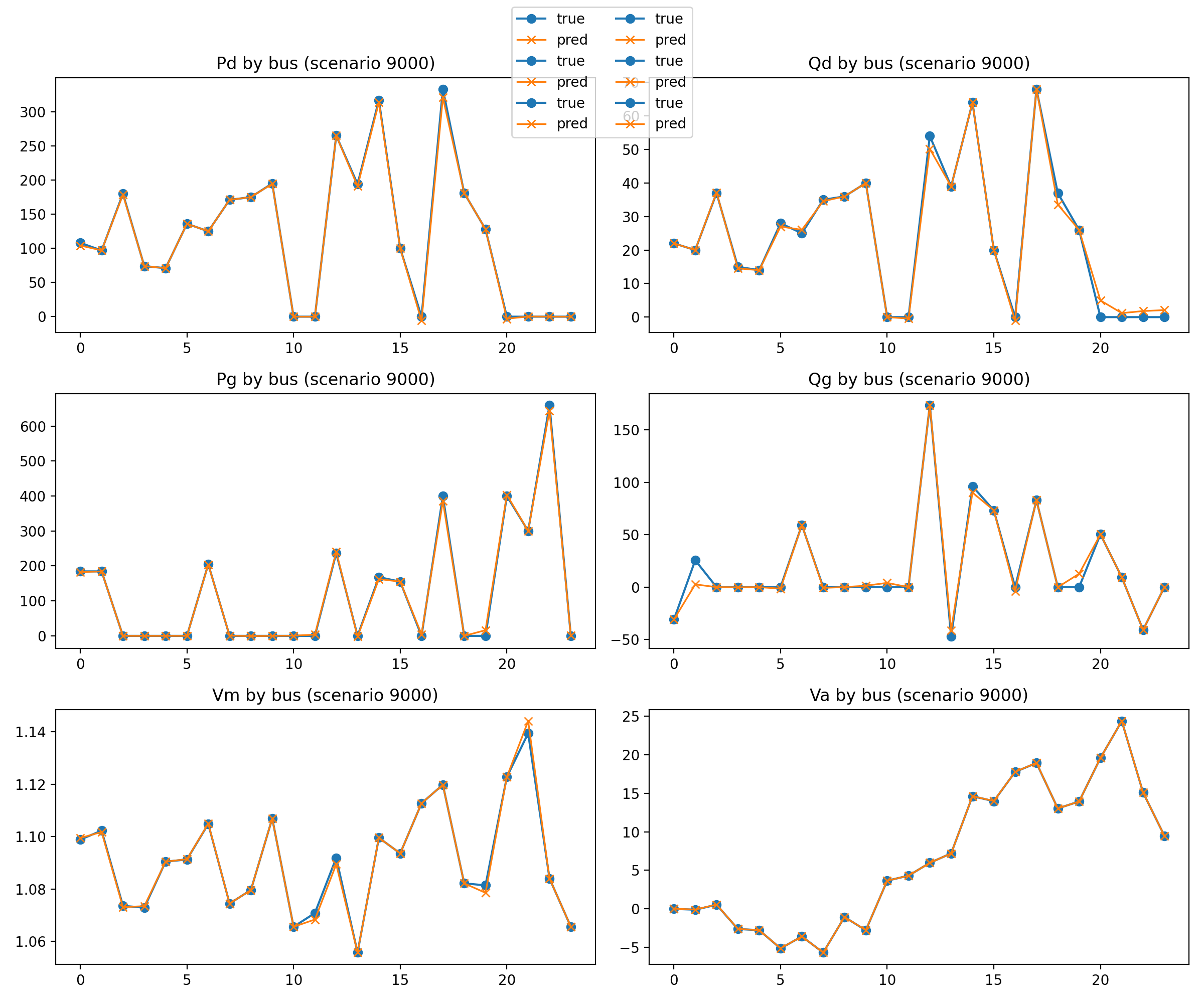}
  \caption{Per-bus true (blue) and predicted (orange) curves for all six target variables over the 24 buses.}
  \label{fig:perbus}
\end{figure}

Fig.~\ref{fig:perbus} shows per-bus true and predicted curves for all six targets across the 24 buses in a representative test scenario, confirming that the surrogate reproduces the spatial variation of each quantity rather than only matching aggregate statistics.

\subsection{Solver Cross-Validation: Julia/GridFM vs.\ PyPower}
To confirm the mathematical correctness of the Julia/PowerModels.jl droop
implementation independently of the PSCAD comparison in
Section~III-A, the Julia solver is cross-validated against a custom
Python droop solver (PyPower-Droop) configured with identical network
data and control parameters on the IEEE 24-bus RTS.

\subsubsection{Validation Setup}
Both solvers are initialized with identical loads, generator setpoints,
and admittance data. The droop-controlled set comprises buses
$\{1,2,7,13,14,15,16,18,21,22,23\}$ (10 buses, 32 generators), with frequency
droop gains $m_p \in [0.03,0.05]$~p.u.\ (randomized in Julia; fixed at
$0.04$~p.u.\ in PyPower) and voltage droop gains $m_q \in [0.02,0.04]$~p.u.\
(fixed at $0.03$~p.u.\ in PyPower). A frequency deadband of $\pm0.006$~p.u.\ ($\pm0.36$~Hz at 60~Hz) is applied in both. 
The system is initialized with a scheduled generation level that does not cover demand plus transmission losses, so as to elicit a measurable droop response. The Julia solver is evaluated across 10 scenarios with topology and admittance perturbations (156 total scenarios); PyPower is run on the primary reference scenario for direct comparison.

\subsubsection{Frequency Deviation Agreement}
Table~\ref{tab:crossval} compares key steady-state quantities. The two
solvers agree to within $0.000025$~p.u.\ in frequency deviation ($\Delta f$),
a $1.51\%$ relative error. The negative $\Delta f \approx -0.0016$~p.u.\ in
both solvers is physically consistent with scheduled generation being
insufficient to cover demand plus losses: the droop-controlled generators
raise their output in response to the frequency decline, and the system
settles at a new equilibrium near $59.90$~Hz.

\subsubsection{Source of Residual Discrepancy}
The $1.51\%$ difference is fully accounted for by AC transmission-loss
modeling. As shown in Table~\ref{tab:crossval}, PyPower reports
$47.22$~MW of losses versus $36.61$~MW in Julia, a $10.61$~MW gap. This
larger loss component demands a proportionally larger droop response,
shifting the equilibrium frequency. Crucially, the droop \emph{response}
itself---the mapping from power imbalance to frequency deviation---is
reproduced consistently across both solvers, confirming that the
discrepancy stems from a platform-level loss-modeling choice rather than
an error in the droop implementation.

\subsubsection{Validation Summary}
The study confirms three properties of the Julia/GridFM droop
implementation: \emph{mathematical correctness}, both solvers reach
near-identical frequency deviations from the same base network and
control parameters, verifying the droop equality constraints within the
AC power-flow problem; \emph{physical consistency}, the system exhibits
the expected frequency drop under generation deficit, converging within
the bound set by platform-level loss differences; and \emph{robustness},
the solver converges to a stable operating point across all 156
scenarios. Together with the PSCAD validation of Section~III-A, these
results establish the Julia/PowerModels.jl droop formulation as a
numerically accurate data-generation engine for the GNN training
datasets.

\subsection{Control Response Analysis}
To characterize the behavior of the three operational modes, two
complementary scalar metrics are computed per scenario. These metrics
quantify the \emph{physical} control response of the droop fleet; they
characterize what the deadband does to the system, and are distinct from
the surrogate-accuracy metrics of Section~III-B.

The control error $a_c$ quantifies the relative suppression of the droop response introduced by the deadband:
 
\begin{equation}
a_c = \frac{|\Delta f_{\mathrm{eff}} - \Delta f|}{|\Delta f| + \epsilon_0},
\label{eq:ac}
\end{equation}

where $\Delta f$ is the raw frequency deviation, $\Delta f_{\mathrm{eff}}$ is the deadband-processed deviation from~(3), and   $\epsilon_0 = 10^{-6}$~p.u.\ prevents division by zero. When the deadband is inactive ($\Delta f_{\mathrm{eff}} \to \Delta f$), $a_c \to 0$; when it fully suppresses the response ($\Delta f_{\mathrm{eff}} \approx \epsilon\, \Delta f$), $a_c \to 1-\epsilon \approx 0.95$. Values $a_c > 1$ arise when the smooth activation~(4) produces a $\Delta f_{\mathrm{eff}}$ that overshoots $\Delta f$ near the deadband boundary, corresponding to the large outliers seen in the deadband mode.

The droop compensation $d_c$ measures the total normalized corrective active power delivered by the droop fleet:
 
\begin{equation}
d_c = \frac{1}{S_{\mathrm{base}}\,|\mathcal{D}|}
\sum_{i \in \mathcal{D}} |P_{g,i} - P^{\mathrm{set}}_{g,i}|,
\label{eq:dc}
\end{equation}

where $S_{\mathrm{base}}$ is the system MVA base and $|\mathcal{D}|$ is the number of droop-participating generators. Normalizing by both base and fleet size makes $d_c$ comparable across system scales.

Table~\ref{tab:control} reports the mean and standard deviation of both
metrics across all scenarios in each mode. Adding the frequency deadband raises mean $a_c$ from $0.083$ (droop only) to $0.225$, with the standard deviation widening from $0.052$ to $0.236$: by suppressing the droop response for deviations below the deadband half-width ($db = 0.006$~p.u.), the controller intentionally leaves small deviations uncorrected. This wider spread is not a performance degradation---it reflects the band within which deviations are deliberately tolerated to reduce unnecessary actuation and the associated generator cycling.

\begin{table}[t]
\centering
\caption{Control Response Metrics Across Operational Modes:
Control Error ($a_c$) and Droop Compensation ($d_c$)}
\label{tab:control}
\begin{tabular}{lccc}
\hline
 & Droop & Droop & Traditional \\
Metric & w/ Deadband & w/o Deadband & (Slack) \\
\hline
Mean $a_c$ & 0.2245  & 0.0828  & 0.0137 \\
Std $a_c$  & 0.2361  & 0.0515  & 0.0067 \\
Mean $d_c$ & 0.00238 & 0.00219 & 0.00252 \\
Std $d_c$  & 0.00065 & 0.00070 & 0.00072 \\
\hline
\end{tabular}
\end{table}

By contrast, mean $d_c$ is statistically consistent across all three modes ($0.00238$, $0.00219$, $0.00252$; within $8\%$), with even closer standard deviations. The total corrective power delivered by the droop fleet is therefore invariant to the deadband configuration: the deadband governs \emph{when} generators respond, not the total energy correction. The deadband thus introduces a well-defined trade-off---control error increases and widens, while total compensation is unchanged---making the deadband width a practical design parameter for operators seeking to reduce generator cycling without compromising power balance. This behavioral distinctiveness is precisely what motivates exposing the deadband as a node feature in the dataset.

\begin{table*}[ht]
\centering
\caption{GNN Surrogate Test-Set Performance on the IEEE 300-Bus System,
Stratified by Bus Type}
\label{tab:gnn300}
\small
\begin{tabular}{llcccccc}
\hline
Bus Type & Metric & $P_d$ (MW) & $Q_d$ (MVAr) & $P_g$ (MW) & $Q_g$ (MVAr) & $V_m$ (p.u.) & $V_a$ (deg) \\
\hline
\multirow{4}{*}{PQ (231)}
 & RMSE & 7.50 & 5.58 & --\textsuperscript{a} & --\textsuperscript{a} & 0.0039 & 0.1202 \\
 & MAE  & 3.89 & 2.64 & --\textsuperscript{a} & --\textsuperscript{a} & 0.0020 & 0.0699 \\
 & Bias & 0.11 & $-0.77$ & --\textsuperscript{a} & --\textsuperscript{a} & 0.0010 & 0.0609 \\
 & $R^2$ & 0.9998 & 0.9994 & --\textsuperscript{a} & --\textsuperscript{a} & 0.9802 & 0.9998 \\
\hline
\multirow{4}{*}{PV (68)}
 & RMSE & 11.36 & 7.89 & 11.30 & 8.45 & 0.0022 & 0.1237 \\
 & MAE  & 8.96 & 5.30 & 8.28 & 6.72 & 0.0014 & 0.1015 \\
 & Bias & $-4.34$ & 2.68 & 6.40 & 6.27 & $-0.0010$ & 0.1008 \\
 & $R^2$ & 0.9998 & 0.9992 & 1.0000 & 0.9998 & 0.9931 & 0.9999 \\
\hline
\multirow{4}{*}{REF (1)}
 & RMSE & 4.98 & 12.64 & 11.45 & 6.58 & 0.0022 & --\textsuperscript{b} \\
 & MAE  & 4.19 & 11.90 & 8.32 & 5.39 & 0.0021 & --\textsuperscript{b} \\
 & Bias & $-3.88$ & $-11.76$ & $-5.32$ & $-3.68$ & $-0.0020$ & --\textsuperscript{b} \\
 & $R^2$ & --\textsuperscript{b} & --\textsuperscript{b} & 0.9946 & --\textsuperscript{c} & 0.6694 & --\textsuperscript{b} \\
\hline
\multirow{4}{*}{System (PQ+PV)}
 & RMSE & 8.54 & 6.18 & 7.04 & 5.43 & 0.0036 & 0.1210 \\
 & MAE  & 5.04 & 3.25 & 5.08 & 4.15 & 0.0019 & 0.0771 \\
 & Bias & $-0.90$ & 0.01 & $-0.97$ & 3.58 & 0.0005 & 0.0700 \\
 & $R^2$ & 0.9998 & 0.9994 & 1.0000 & 0.9997 & 0.9841 & 0.9998 \\
\hline
\end{tabular}
\begin{flushleft}\footnotesize
\textsuperscript{a} PQ buses carry no generation ($P_g = Q_g \equiv 0$); error and $R^2$ are not meaningful for a constant-zero target.\\
\textsuperscript{b} The reference bus fixes $V_a = 0^\circ$ and demand at constant values across scenarios; $R^2$ is degenerate for near-constant targets and is omitted.\\
\textsuperscript{c} REF-bus $Q_g$ yields $R^2 = -1.2\times10^{8}$ (near-constant target, large relative variance); reported as degenerate and omitted.
\end{flushleft}
\end{table*}

\subsection{Scalability: IEEE 300-Bus System}
\label{sec:300bus}

A dataset of 9{,}994 converged operating scenarios is generated for the
IEEE 300-bus system following the protocol of Section~II, assigning droop gains and frequency/voltage deadband parameters to the 68 droop-participating PV buses and the reference bus as node features. The dataset is split into 7{,}996 training, 999 validation, and 999 test scenarios, and the surrogate is trained via masked reconstruction with a physics-informed loss using the same architecture and optimizer configuration as the 24-bus case. The data-generation solve converged for effectively all sampled scenarios ($9{,}994$ retained), corresponding to a negligible discard rate; the distributed droop mechanism prevents the voltage-collapse divergences that would otherwise remove stressed operating points, so the retained set is not biased toward well-conditioned cases.

Table~\ref{tab:gnn300} reports test-set performance over 299{,}700
bus-level samples, stratified by bus type. The system-wide surrogate
achieves $R^2 = 1.0000$ for active generation, $R^2 = 0.9998/0.9994$ for
active/reactive load, and $R^2 = 0.9841$ for voltage magnitude (RMSE $= 0.0036$~p.u.), comparable to the 24-bus results (Table \ref{tab:gnn_24bus}) and confirming scalability across a network with substantially greater topological and impedance heterogeneity. Reactive quantities exhibit the largest absolute errors at PV buses (RMSE $7.89$--$8.45$~MVAr), consistent with the sensitivity of reactive injections to local voltage under the nonlinear $Q$--$V$ characteristic observed on the 24-bus system. Degenerate entries at the single reference bus ($V_a$ fixed, near-constant demand) yield unstable $R^2$ for near-constant targets and are omitted. All 299{,}700 predictions remain within the physical voltage range $[0.9, 1.1]$~p.u., confirming that the droop-augmented model preserves grid feasibility at scale.


\section{Conclusion}
\label{sec:conclusion}
This paper presented a droop-aware extension of the GridFM framework, incorporating $P$--$f$ and $Q$--$V$ droop control with smooth deadband modeling as per-bus node features to enable control-aware AC power-flow analysis. The deadband discontinuity was resolved via a $\tanh$ approximation that preserves solver differentiability, allowing deadband width to be treated as a differentiable node feature. The implementation was validated against a PSCAD electromagnetic-transient reference ($0.11\%$ maximum steady-state error) and cross-validated against an independent PyPower droop solver ($1.51\%$ relative frequency deviation, principally attributable to loss-model differences). A transformer-based GNN trained on the droop-augmented dataset achieved $R^2 = 0.9996$ for active power generation and $0.0015$~p.u.\ voltage-magnitude RMSE on the IEEE 24-bus RTS, with scalability confirmed on the IEEE 300-bus system across $299{,}700$ test predictions, all within physical voltage limits. A three-mode control study further showed that the frequency deadband widens the control-error distribution while leaving total droop compensation unchanged, establishing deadband width as an actionable design parameter for reducing generator cycling without compromising power balance. 

Collectively, these results establish droop gain and deadband width as actionable node-level features for GNN power-flow surrogates, enabling rapid scenario sweeps over control configurations as an alternative to repeated numerical solves. Future work will quantify inference speedup against classical AC power-flow solvers, benchmark the pre-trained surrogate against task-specific and from-scratch baselines to isolate the contribution of foundation-model pre-training, and extend the framework toward cross-topology generalization and additional control-aware downstream tasks.

\bibliographystyle{IEEEtran} 
 \bibliography{Paper1}

\end{document}